\documentclass[final]{ustcstep}

\usepackage{verbatim}
\usepackage{graphicx}
\usepackage{lscape}
\usepackage{multirow}
\usepackage{latexsym}

\usepackage{color}

\newcommand{\mod}[1]{#1}

\newcommand\addr[2]{{\footnotesize \it $^{#1}$#2}\\}

\begin{document}

\title{Super-elastic Collision of Large-scale Magnetized Plasmoids in The Heliosphere}


\author{Chenglong Shen$^{1}$, Yuming Wang$^{1,*}$, Shui Wang$^1$, Ying Liu$^{2,4}$, Rui Liu$^1$, Angelos Vourlidas$^3$, \\[1pt]
Bin Miao$^1$, Pinzhong Ye$^1$, Jiajia Liu$^1$, and Zhenjun Zhou$^{1}$\\[1pt]
\addr{1}{CAS Key Laboratory of Geospace Environment, Department of Geophysics and Planetary Sciences, University of Science \&}
\addr{ }{ Technology of China, Hefei 230026, China}
\addr{2}{Space Sciences Laboratory, University of California, Berkeley, CA 94720, USA}
\addr{3}{Space Science Division, Naval Research Laboratory, Washington, D.C. 20375, USA}
\addr{4}{State Key Laboratory of Space Weather, National Space Science Center, Chinese Academy of Sciences, Beijing, China}
\addr{*}{Corresponding author. E-mail: ymwang@ustc.edu.cn}}

\maketitle

\begin{abstract}
Super-elastic collision is an abnormal collisional process, in which
some particular mechanisms cause the kinetic energy of the system
increasing. Most studies in this aspect focus on solid-like objects,
but they rarely consider gases or liquids, as the collision of the
latter is primarily a mixing process. With cross-field diffusion
being effectively prohibited, magnetized plasmoids are different
from ordinary gases. But it remains unclear how they act during a
collision. Here we present the global picture of a unique collision
between two coronal mass ejections in the heliosphere, which are the
largest magnetized plasmoids erupting from the Sun. Our analysis for
the first time reveals that these two magnetized plasmoids collided
like solid-like objects with a \mod{73\%} likelihood of being
super-elastic. Their total kinetic energy surprisingly increased by
about \mod{6.6\%} through the collision, which significantly
influenced the dynamics of the plasmoids.
\end{abstract}

\section{Introduction}
Collisional dynamics is essential in determining global structure
and evolution of macro- and micro- objects, like planet
rings\cite{Bridges_etal_1984}, granular
materials\cite{Louge_Adams_2002}, and
nanoclusters\cite{Kuninaka_Hayakawa_2009, Saitoh_etal_2010}. To
classify collisions in terms of energy transfer, Newton defined the
coefficient of restitution, $e$, which is normally between $0$ and
$1$. However, abnormal $e$ values, such as $e > 1$ (ref.
\cite{Louge_Adams_2002, Kuninaka_Hayakawa_2004, Smith_Liu_1992,
Calsamiglia_etal_1999}) or $e<0$ (ref. \cite{Saitoh_etal_2010}) have
been reported. A super-elastic collision is a process through
which the linear kinetic energy of the collisional system increases,
i.e., $|e|>1$. In the literature, there have been several
mechanisms proposed to explain such an abnormal increase of linear
kinetic energy during a collision. In granular physics, for example,
the oblique impact collision with local deformation may help
transfer rotational kinetic energy into linear kinetic
energy\cite{Louge_Adams_2002, Kuninaka_Hayakawa_2004,
Saitoh_etal_2010} (hereafter kinetic energy refers to linear kinetic
energy). Thermal fluctuations are suggested as another possible
reason leading to super-elastic collisions of
nanoclusters\cite{Kuninaka_Hayakawa_2009}.


In absence of internal magnetic fields, two encountering plasmoids
tend to mix together, just like ordinary gases. But it is unclear
what would happen if they carry strong magnetic fields, especially
in regards to the nature of collision and the energy exchange
between them. Coronal mass ejections (CMEs) are
large-scale\cite{Lepping_etal_1990} magnetized plasmoids,
originating from the solar atmosphere and expanding and propagating
into the heliosphere. Since they are a frequently-occurring
phenomenon with an occurrence rate of 4 -- 5 CMEs per day during
solar maximum\cite{Yashiro_etal_2004}, the encounters and
interactions between CMEs are unavoidable. Actually, as a
consequence of interactions, multiple-interplanetary-CME structures
are often observed by in situ instruments\cite{Burlaga_etal_2002,
Wang_etal_2002b, Wang_etal_2003c, Wang_etal_2003a,
Farrugia_Berdichevsky_2004}. Thus the issue of magnetized plasmoid
collision may be addressed by investigating observations of CMEs.

However, the CME dynamics in the heliosphere constitute an intricate
problem\cite{Wang_etal_2006a, Harrison_etal_2009, Liu_etal_2010a,
Liu_etal_2011}, especially when the collision/interaction between
CMEs is involved\cite{Gopalswamy_etal_2001c, Wang_etal_2002b,
Shen_etal_2008, Lugaz_etal_2009, Liu_etal_2012}. The dynamics of two
successive CMEs of 24 -- 25 January 2007 was discussed by Lugaz {\it
et al}.\cite{Lugaz_etal_2009}. They proposed four different
scenarios to explain observations, one of which they think is a
mysterious collision through which the leading CME gained momentum
and finally became faster than the overtaking CME. Most recently, a
CME-CME interaction event of 1 August 2010 has been intensively
studied with a focus on the CME dynamics, CME-driven shock and radio
bursts\cite{Liu_etal_2012, MartinezOliveros_etal_2012,
Temmer_etal_2012}.
Numerical simulations of the interaction between CMEs were also
carried out by many researchers\cite{Gonzalez-Esparza_etal_2004,
Schmidt_Cargill_2004, Wang_etal_2005, Lugaz_etal_2005,
Hayashi_etal_2006, Xiong_etal_2007, Xiong_etal_2009,
Shen_etal_2011a}, but few discussed the nature of the CME
collisions.

During 2 -- 8 November 2008, the Sun Earth Connection Coronal and Heliospheric Investigation
(SECCHI) suites\cite{Howard_etal_2008} onboard the twin Solar TErrestrial
RElations Observatories (STEREO)\cite{Kaiser_etal_2008}
 captured the process of the
chasing and colliding of two CMEs in the heliosphere with clear
imaging observations. Each SECCHI suite carries the
cameras COR1, COR2, HI1 and HI2, and can seamlessly track
CMEs from the corona to interplanetary space.
Since the events occurred near the solar
minimum, the conditions in the heliosphere were quite simple. The
events provide us with a unique opportunity to study the physical details
of CME collisions. As will be seen, the collision between the two
CMEs is super-elastic in nature, during which their
total kinetic energy increased. These results advance our
understanding of the behavior of large-scale magnetized plasmoids.

\section{Imaging of two successive CMEs and their collision}
\begin{figure*}
\includegraphics[width=\hsize]{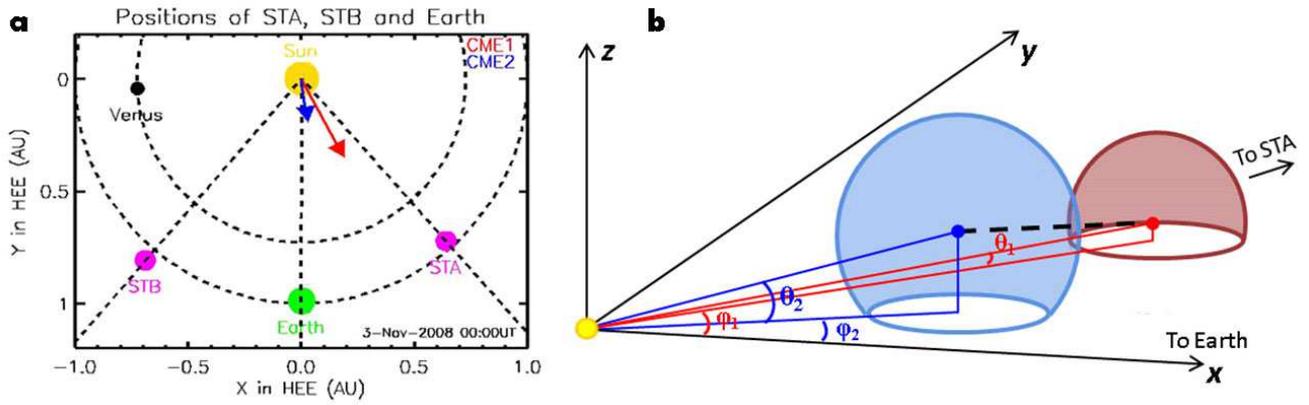}
\caption{The sketch map of (a) the positions of spacecraft and (b) the collision of the CMEs.}\label{sketch}
\end{figure*}

The two CMEs originated from the Sun at about 00:35 UT and 22:35 UT,
respectively, on 2 November 2008, when STEREO-A spacecraft was
located at 0.97 AU and 41$^\circ$ to the west of the Sun-Earth line,
while STEREO-B was located at 1.07 AU and 40$^\circ$ to the east
(Fig.\ref{sketch}a). These events were reported by Kilpua
{\it et al.}\cite{Kilpua_etal_2009} with a focus on their solar
source locations and in situ effects at 1 AU. One can refer to that
paper or Sec.2 of Supplementary Information for the details of the
propagation of the two CMEs in the corona. Here we focus on their
collision in the heliosphere.

\begin{figure*}
\begin{center}
\includegraphics[width=\hsize]{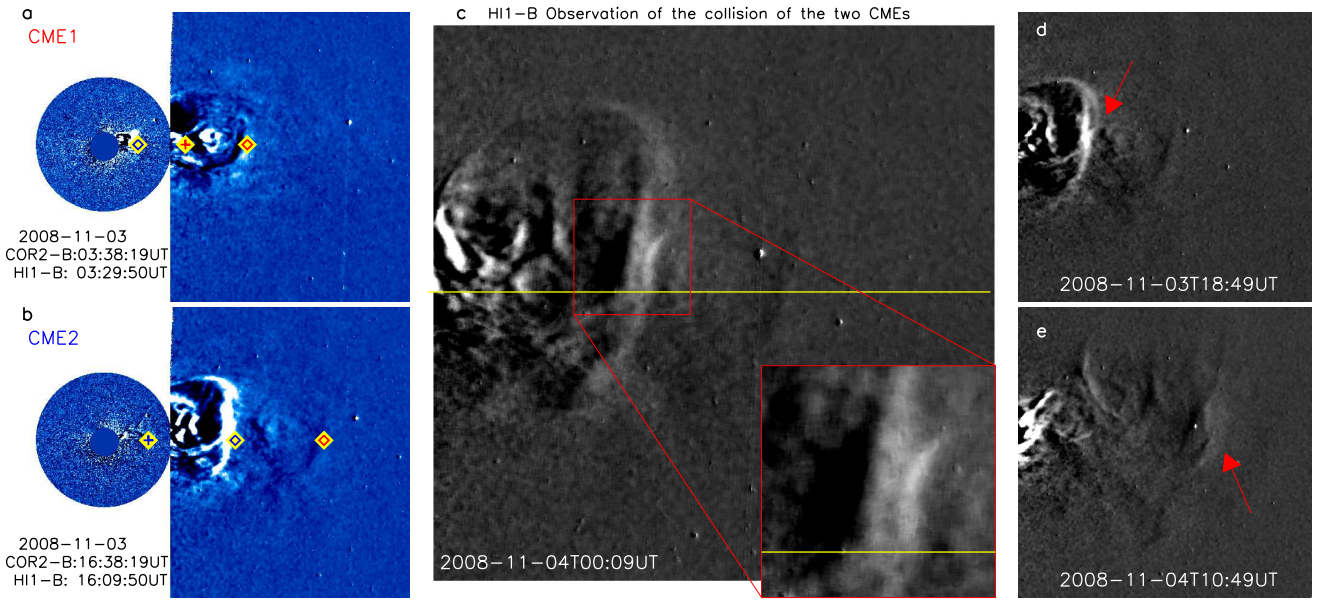}
\caption{The STEREO/SECCHI images of the two CMEs and their
collision in the heliosphere. (a) and (b) are the running-difference
images showing CME1 and CME2. The red symbol `$\Diamond$' and `$+$'
mark the front and rear edges of CME1, respectively, and the blue
symbols for CME2. (c) The running-difference image of HI1-B showing
the collision of the two CMEs. (d) and (e) show the beginning and
end of the collision, and the red arrows indicate the collision
region.} \label{overview}
\end{center}
\end{figure*}

Being faster than CME1, CME2 finally caught up and collided with
CME1. This phenomenon was clearly recorded by HI1 onboard STEREO-B,
or HI1-B briefly. Based on the HI1-B images, we can see that the
distance between the front edge of CME2 and the rear edge of CME1
became smaller and smaller. The apparent touch of the two CMEs began
approximately around 18:49 UT on 3 November 2008, which was
registered as a significant enhancement of brightness around an
arc-shaped structure (Fig.\ref{overview}d). We call the
brightness enhanced region as collision region, and the arc
structure is the core of the region. Since the arc structure is
caving into CME2, the brightness enhancement is not simply due to
superposition of the two CMEs, but probably the result of a soft
object colliding with a hard object. Actually, if the two CMEs did
not collide, the kinetic evolution of CME1 cannot be explained only
by solar wind acceleration (refer to Sec.11 of Supplementary
Information). The brightened arc structure stayed visible for about
7 hours with the most clear appearance at around 00:09 UT on 4
November (Fig.\ref{overview}c).
\mod{The whole collision region remained brightened
much longer} till 10:49 UT on 4 November 2008
(Fig.\ref{overview}e). It seems that the entire collisional process
of such large-scale magnetized plasmoids is similar to that of
elastic balls, which includes a pre-collision phase, a compression
phase, a restitution phase and a post-collision phase. We think that
the appearance and disappearance of the visible arc structure define
the start and the end of the compression phase, respectively, and
the complete disappearance of the brightened region between the two
CMEs marks the end of the restitution phase, i.e., the end of the
collision between them.
The movies are available as
online materials.

\section{Tracking and Dynamics of the two CMEs in the heliosphere}
\begin{figure*}
\includegraphics[width=\hsize]{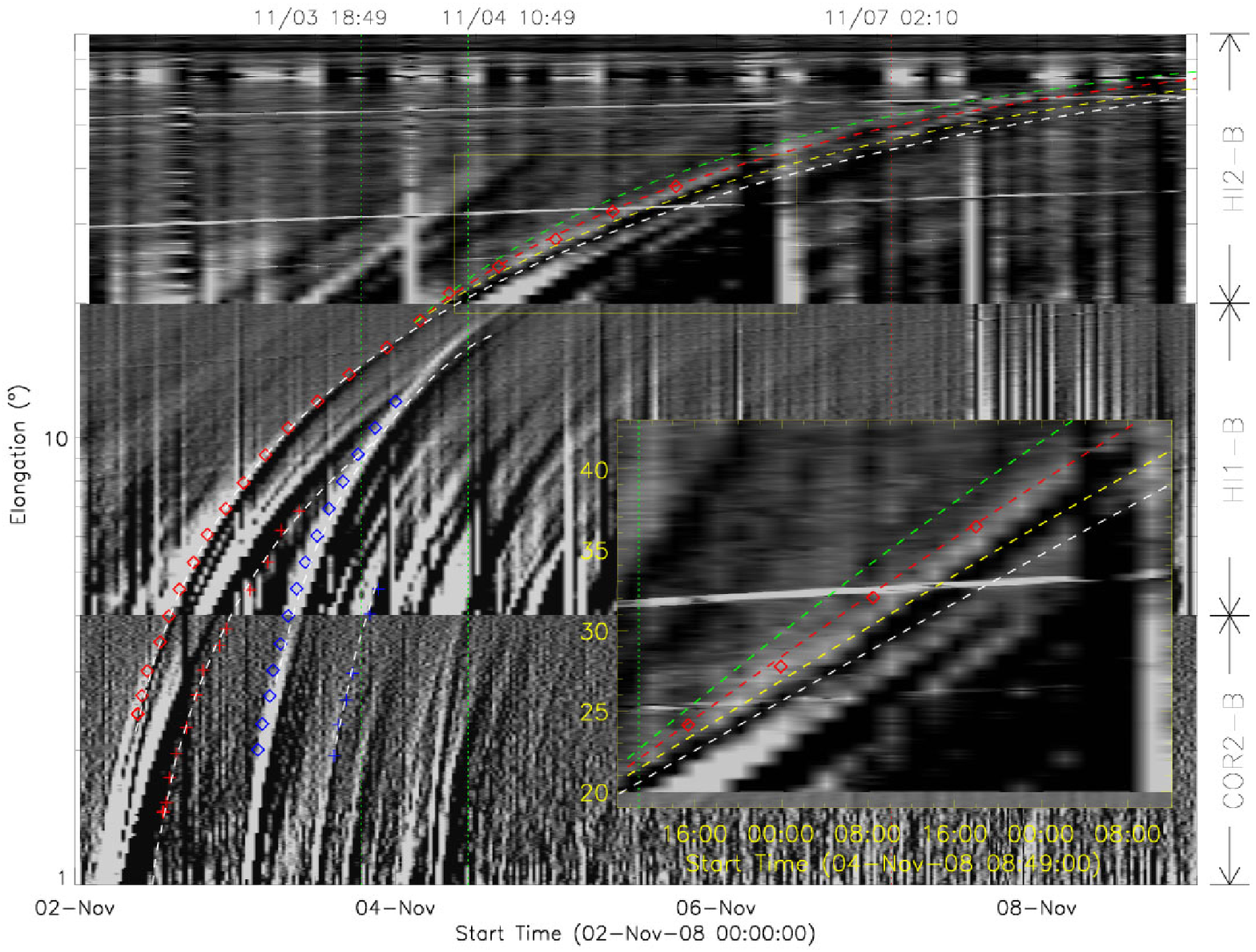}
\caption{The time-elongation map from 2 to 9 November 2008
constructed based on the running-difference images from STEREO-B.
The symbol of `$\Diamond$' and `$+$' show the front and rear edges
of the CMEs, respectively. The two vertical dotted green lines
indicate the start and end of the collision. The red vertical line
marks the arrival time of CME1 at STEREO-A.
The region enclosed by yellow rectangle is zoomed-in in the lower-right corner. Refer to the
main text for more details.} \label{jmap}
\end{figure*}

In order to analyze the dynamics of the CMEs and their collision, a
time-elongation map, known as Jmap\cite{Sheeley_etal_1999,
Davies_etal_2009, Liu_etal_2010a, Liu_etal_2010}, is constructed. To
facilitate the comparison between imaging data and in situ data at 1
AU, a 64-pixel wide slice is placed along the ecliptic plane in the
running-difference images from COR2, HI1 and HI2 onboard STEREO-B to
produce the Jmap (Fig.\ref{jmap}).
A bright-dark alternating track from lower-left to upper-right
usually indicates a bright structure moving away from the Sun. The
two vertical dotted green lines mark the start and end times of the
collision.

The front edges of CME1 and CME2 are distinct in the Jmap as marked
by the red and blue `$\Diamond$', respectively. They are the same
points marked by the red and blue `$\Diamond$' in
Fig.\ref{overview}a and \ref{overview}b. The rear edges of the two
CMEs are not clear in the Jmap. To find out where the tracks of the
rear edges of the two CMEs are, we directly identify their rear
edges in coronagraph images like in Fig.\ref{overview}a and
\ref{overview}b, and then dot them back to the Jmap as shown by the
red and blue `$+$', respectively.
Note that the significant track between the red
`$\Diamond$' and `$+$' symbols does not correspond to the CME1's
rear edge but to its bright core.

The elongation angle of a given feature in the Jmap can be converted
to the heliocentric distance under some assumptions\cite{Lugaz_2010,
Lugaz_etal_2010, Liu_etal_2010a, Liu_etal_2010}. An often used
assumption is to approximate a CME as a sphere\cite{Lugaz_etal_2009,
Shen_etal_2011a}. By further assuming that the front and rear edges
recorded in the Jmap are the points of tangency determined by the
circular cross-section of the CME in the ecliptic plane and the
observer STEREO-B, we get the heliocentric distance of the CME
center, $d$, its radius $r$, and their projected components on the
ecliptic plane, $d_p$ and $r_p$, in terms of the heliocentric
distance, $l$, of STEREO-B, the elongation angles, $\varepsilon_F$
and $\varepsilon_R$, of the CME front and rear edges, and the
latitude, $\theta$, and longitude, $\varphi$, of the CME center. The
detailed derivation can be found in the Sec.4 of Supplementary
Information. Due to the presence of the solar wind stretching
effect, a CME might become `pancake' shaped even if it was initially
spherical\cite{Riley_Crooker_2004, Liu_etal_2006a}. \mod{The HI1
imaging data suggests that the effect is somewhat significant for CME2,
but not for CME1. Thus a small correction is made to CME2 to reduce the
effect (refer to Sec.5 of Supplementary Information).}

With the aid of the Graduated Cylindrical Shell (GCS)
model\cite{Thernisien_etal_2006, Thernisien_2011}, the latitude,
$\theta$, and longitude, $\varphi$, of the two CME centers can be
obtained from COR2 images. It is found that both CMEs propagated
almost radially with a nearly constant longitude and latitude in the
COR2 FOV\cite{Kilpua_etal_2009}, which are listed in
Table~\ref{table} (refer to Sec.2 and 3 of Supplementary
Information for details). As the interplanetary magnetic field and solar wind
density get weaker and lower, respectively, farther away from the
Sun, it is reasonable to assume that they would keep their
propagation directions in the HI1 FOV until the collision. The
results given by the model suggest that both CMEs propagated between
the Sun-Earth line and the Sun-STEREO-A line with CME1 closer to the
latter line and CME2 closer to the former, which is in agreement with
the previous study\cite{Kilpua_etal_2009}.

\begin{table*}
\centering
\caption{The parameters of the two CMEs before and after the collision.}\label{table}
\tabcolsep 2.3pt
\begin{tabular}{c|ccccccccccccccc}
\hline
\hline
\multicolumn{16}{c}{Parameters derived from observations}\\
\hline
   &\multicolumn{3}{c}{$\theta$} &  \multicolumn{3}{c}{$\varphi$} & \multicolumn{3}{c}{$v_{c}$}  & \multicolumn{3}{c}{$v_{e}$}\\
\hline
CME1  &    \multicolumn{3}{c}{6$\pm2$}  &   \multicolumn{3}{c}{28$\pm10$}  & \multicolumn{3}{c}{\mod{$243^{+25}_{-16}$}} &  \multicolumn{3}{c}{ \mod{$43^{+16}_{-15}$}} \\
CME2  &   \multicolumn{3}{c}{16$\pm2$}  &   \multicolumn{3}{c}{ 8$\pm10$}  & \multicolumn{3}{c}{\mod{$407^{+102}_{-74}$}} & \multicolumn{3}{c}{74\mod{$^{+65}_{-51}$}}\\
\hline
\hline
\multicolumn{16}{c}{Second-level derived parameters} \\
\hline
   &$v_{p}$ & $v_{ep}$ & $\theta_C$ & $\varphi_C$ &$v_\perp$ & $v_\parallel$ & $v^\prime_\parallel$ & $v^\prime_c$ & $v^\prime_p$ &$v^\prime_{ep}$ & $\Delta \theta_v$ & $\Delta \varphi_v$ & $\Delta$E/E & $\Delta$E$_t$/E$_t$ & $e$\\
 \hline
CME1  &    \mod{241} & \mod{36}&  \multirow{2}{*}{\mod{-10}} &   \multirow{2}{*}{\mod{57}}  &  130  &   \mod{205}  &  \mod{288}  & \mod{316} & \mod{316} & \mod{41} & -4  & 7  &  68\%  & \multirow{2}{*}{\mod{6.6\%}} & \multirow{2}{*}{\mod{5.4}} \\
CME2  &   \mod{392} & \mod{26} &   &   &  \mod{332}  &   \mod{237}  &  \mod{116}  & \mod{351} & \mod{325}  & N/A$^*$&  \mod{6}  &  \mod{-16}  & \mod{-25\%}  & & \\
\hline
\hline
\end{tabular}\\
\footnotesize
$\theta$ and $\varphi$ are the CME's latitude and longitude. $v_c$ and $v_e$ are the
propagation and expansion speed of a CME, derived from the Jmap by assuming the CME
is a sphere (refer to Sec.4 and 5 of Supplementary Information). $v_{p}$ and $v_{ep}$ are the average values of
the components of $v_c$ and $v_e$ in the ecliptic plane, respectively. $\theta_C$ and $\varphi_C$ are the
latitude and longitude of the collision direction (refer to Fig.7 in Sec.7 of Supplementary Information).
$v_\perp$ and $v_\parallel$ are the components of the CME velocity perpendicular and
parallel to the collision direction, respectively. The superscript of prime denotes
the parameters after the collision. $\Delta\theta_v$ and
$\Delta\varphi_v$ are the change of the CME velocity.
$\Delta E/E=(E^\prime-E)/E$ is the percentage of the kinetic energy changed, and $E_t$
is the sum of the kinetic energy of the two CMEs. All the angles in the table are in
units of degree, and all the speeds are in units of km s$^{-1}$. Here, only the uncertainties
of $\theta$, $\varphi$, $v_c$ and $v_e$ are listed, \mod{and the uncertainties of speeds have included
the uncertainties in the CMEs' directions}. The uncertainties of the second-level derived
parameters are not listed, but all taken into account in our analysis. $^*$ After the collision,
CME2 left the ecliptic plane, and thus there is no available component of expansion speed in
the ecliptic plane.
\end{table*}

\begin{figure*}
\includegraphics[width=\hsize]{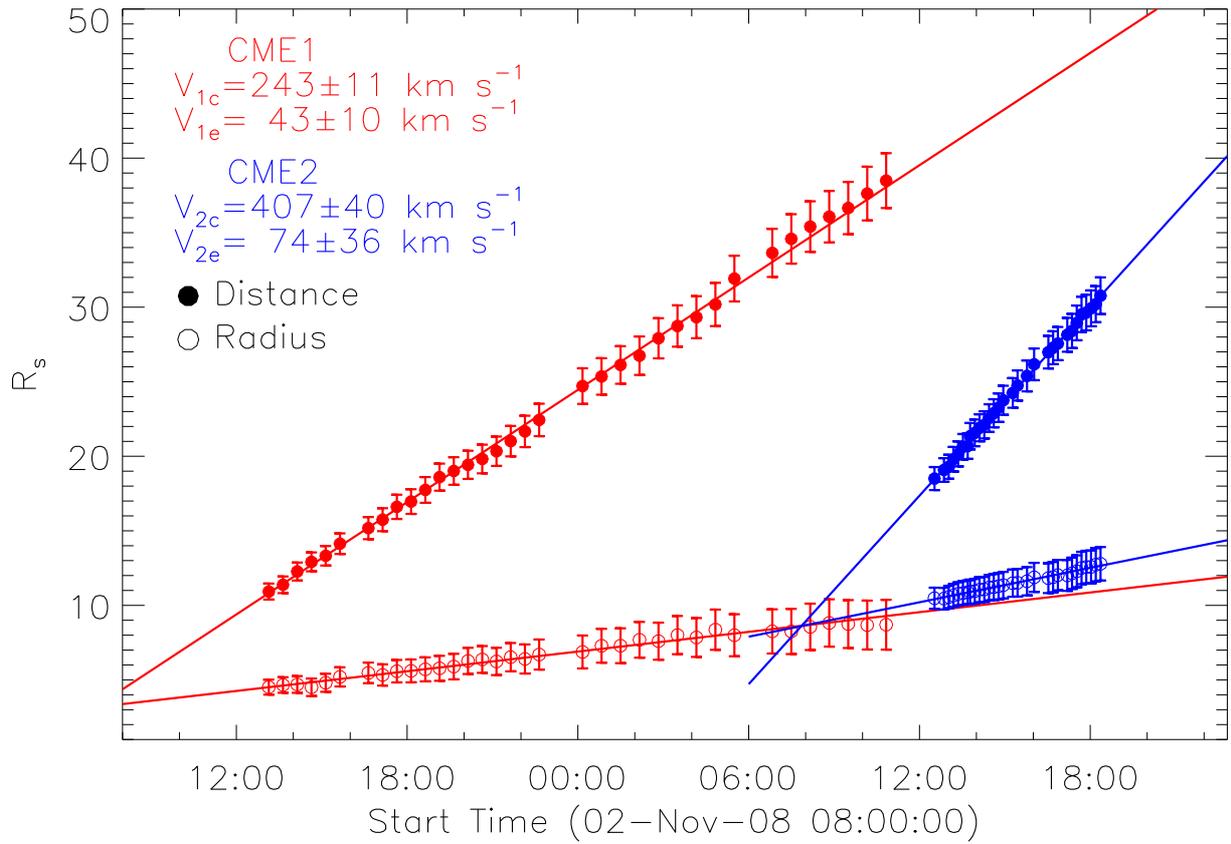}
\caption{The heliocentric distance $d$ (dots) and the radius $r$ (open circles) of the two
CMEs as functions of time \mod{for the case that $\theta$ and $\varphi$ of CME1
are $6^\circ$ and $28^\circ$, respectively and those of CME2 are $16^\circ$
and $8^\circ$, respectively. The error bars are derived from the 5\% uncertainty in
elongation angle. Speeds are obtained by linear fitting with the error bars taken
taken into account, and a 2-$\sigma$ uncertainty is chosen, which makes
the confidence level greater than 95\%.}} \label{speed}
\end{figure*}

Figure~\ref{speed} shows $d$ and $r$ as a function of time for both
CMEs. Since the front and rear edges of the CMEs are more or
less diffused, a reasonable error of $\pm5\%$ in determination of \mod{the elongation
angle of} the CME front and rear edges is considered. \mod{The resultant undertainties of $d$
and $r$ are indicated by the error bars in Figure~\ref{speed}.}
By applying the linear fitting
to $d$ and $r$ \mod{with these uncertainties taken into account}, we get the propagation speed $v_c$ and expansion
speed $v_e$ of the two CMEs, as well as their components in the
ecliptic plane, $v_p$ and $v_{ep}$. \mod{A 2-$\sigma$ uncertainty of the speeds derived from the linear fitting
is applied in the following analysis.} The
excellent consistency between the fitting lines and the data points
suggests that the two CMEs experienced a nearly constant-speed
propagation and expansion in the heliosphere before they
encountered, though a very weak acceleration can be seen for
CME1. \mod{It should be noted that the uncertainties of CMEs' directions may cause additional
uncertainties of CMEs' speeds, and therefore the final values of the uncertainties
of CMEs' speeds (see $v_c$ and $v_e$ listed in Table~\ref{table}) are larger than
those given in Figure~\ref{speed}. Besides, although the front edge of CME2
perhaps traveled faster than background solar wind, observations suggest that it
did not drive an evident shock ahead (refer to Sec.12 of Supplementary Information).}

Further, we reversely derive the elongation angle-time curves from
the above results, and plot them on the Jmap in white dashed lines
in Figure~\ref{jmap}. These white dashed lines are also extrapolated
to the post-collision phase. It is found that the fitting lines
match the observed tracks very well before the collision, but begin
to deviate from the tracks since the beginning of the collision
(particularly note the tracks of the two CMEs' front edges). Such
deviations mean that the collision between the two CMEs must have
significantly changed their propagation directions and/or speeds.

As an attempt, we might as well treat the CMEs approximately as a
expanding ball in the collision. The situations of the two CMEs at
the time of touching have been sketched in Figure~\ref{sketch}b. It
is a collision in 3-D space. Since CME1 was originally propagating
along a radial direction with lower latitude than CME2, the
collision should push CME1 closer to the Sun-STEREO-A line in the
ecliptic plane and CME2 further away from the ecliptic plane. Thus,
it is expected that CME1 would be observed in situ by the
instruments onboard STEREO-A while CME2 would be missed by the in
situ instruments which are all located in the ecliptic plane. The in
situ data at 1 AU do suggest that only CME1 was observed as expected
(refer to Sec.6 of Supplementary Information for more
details). Its propagation and expansion speeds at 1 AU were about
\mod{342} and 30 km s$^{-1}$, respectively. The increased propagation
speed is consistent with our conjecture that CME1 was accelerated by
the collision. The expansion speed is very close to that derived
from Jmap. This fact allows us to reasonably assume that the
expansion speed was recovered after the collision for both
CMEs, though the expansion speed may vary greatly during the
collision and CME2 was not locally observed at 1 AU.

\section{Super-elastic collision and the energy exchange}\label{sec_results}
Imagining two expanding elastic balls, not only their collision will
result in the momentum exchange in the direction connecting the
centroids of the two balls (referred to as collision direction
hereafter), but also their continuous expansion may cause their
centroids to separate farther away. We define the approaching speed
as the speed of the centroid of one ball relative to the other in
the collision direction. Under the assumption that the expansion
speeds remained unchanged before and after collision, the collision
should be super-elastic if the sum of the expansion speeds of the
two balls was larger than the approaching speed before the
collision. \mod{Here we first show the results for the case of the CMEs'
parameters given in Table~\ref{table}, and then analyze the influence
of the uncertainties.}

According to the values listed in Table~\ref{table}, we can
derive that the latitude $\theta_C$ and longitude $\varphi_C$ of the
collision direction at the beginning of the collision, i.e., the
elevation angle and azimuthal angle in the heliocentric coordinate
system, are about $-10^\circ$ and $57^\circ$, respectively.
By resolving the propagation velocity vectors into the components
parallel, $v_{\parallel}$, and perpendicular, $v_{\perp}$, to the
collision direction (refer to Sec.7 of Supplementary
Information), we find that the values of $v_{\parallel}$ of the two
CMEs were \mod{205} and \mod{237} km s$^{-1}$, respectively (listed in
Table~\ref{table}), which give an approaching speed of about
\mod{32} km s$^{-1}$. The sum of the expansion speeds of the two
CMEs was about \mod{117} km s$^{-1}$, much larger than the
approaching speed. Hence a super-elastic collision is expected.

The conservation of momentum requires
$m_1v_{1\parallel}+m_2v_{2\parallel}=m_1v^{\prime}_{1\parallel}+m_2v^{\prime}_{2\parallel}$,
where $m_1$ and $m_2$ are the mass of CME1 and CME2, respectively,
and the prime symbol denotes the parameters after the collision.
Here, we approximately treat the collision phase including the
compression and restitution phases as a black box, and adopt
parameters of the two CMEs before (after) the collision for the
first (second) half period of the collision phase. The influence of
this simplification on our final result is not significant (see
Sec.8 of Supplementary Information).

The mass of a CME can be calculated from calibrated coronagraph
images\cite{Vourlidas_etal_2000}. For CME1 and CME2, the derived
masses based on COR2-B observations are about $1.8\times
10^{12}$ kg and $1.2\times 10^{12}$ kg, respectively. The
Thomson scattering and projection effects have been
corrected\cite{Hundhausen_1993, Vourlidas_Howard_2006}. The mass
ratio of CME1 to CME2 is about 1.5.
Hence, for any given coefficient of restitution $e$, i.e.,
$\frac{v^\prime_{2\parallel}-v^\prime_{1\parallel}}{v_{1\parallel}-v_{2\parallel}}$,
the velocities of the two CMEs after the collision can be obtained
(refer to Sec.7 of Supplementary Information) as well as the
expected tracks of the front and rear edges of both the CMEs in the
Jmap. Actually, our calculation suggests that, no matter which value
of the mass ratio we choose, the super-elastic nature of the
collision, which will be seen below, does not change (refer to
Sec.9 of Supplementary Information).

In the Jmap, only the track of the front edge of CME1 is still
identifiable after the collision. Thus we repeatedly adjust the
value of $e$ to find the best match for the observed track. For
the parameters listed in Table 1 (the influence of the uncertainties
will be addressed in the last paragraph of this section), the red
dashed lines starting at the middle of the collision in
Figure~\ref{jmap} shows the best predicted track of the front edge
of CME1, which gives $e=\mod{5.4}$. As a comparison, the tracks for
$e$ equal to 1 and 10 are presented by the yellow and green dashed
lines, respectively. A zoomed-in image in the lower-left corner of
Figure~\ref{jmap} presents the details. Obviously, the tracks
predicted by both the yellow and green dashed lines get worse. $e=1$
indicates a perfect elastic collision, but the yellow line is
obviously lower than the observed track indicated by the red
`$\Diamond$'. The $0<e<1$ tracks predicted by our calculation would
be located even lower.

As summarized in Table~\ref{table}, through the collision, CME1 was
deflected southwestward and its propagation speed increased from \mod{243}
km s$^{-1}$ to about \mod{316} km s$^{-1}$, while CME2 was deflected
northeastward and its speed decreased from \mod{407} km s$^{-1}$ to
\mod{351} km s$^{-1}$. The in situ propagation speed of CME1
was about 40 km s$^{-1}$ larger than the derived post-collision
speed of CME1. It is probably due to the continuous acceleration by
the solar wind. According to the result, the two CMEs were
separating after the collision (refer to Sec.10 of Supplementary
Information for a preliminary discussion). It is worth noting that
CME2 is completely above the ecliptic plane after the collision.
Therefore, it is not surprising that no counterpart of CME2 was
detected by in situ instruments located in the ecliptic plane.
Further, the kinetic energy of CME1 (the contribution from the CME
expansion has been taken into account) is found to increase by about
68\%, while that of CME2 decreased by about \mod{25\%}. As a
whole, the system gained about \mod{6.6\%} kinetic energy during the
collision.

\begin{figure*}
\includegraphics[width=\hsize]{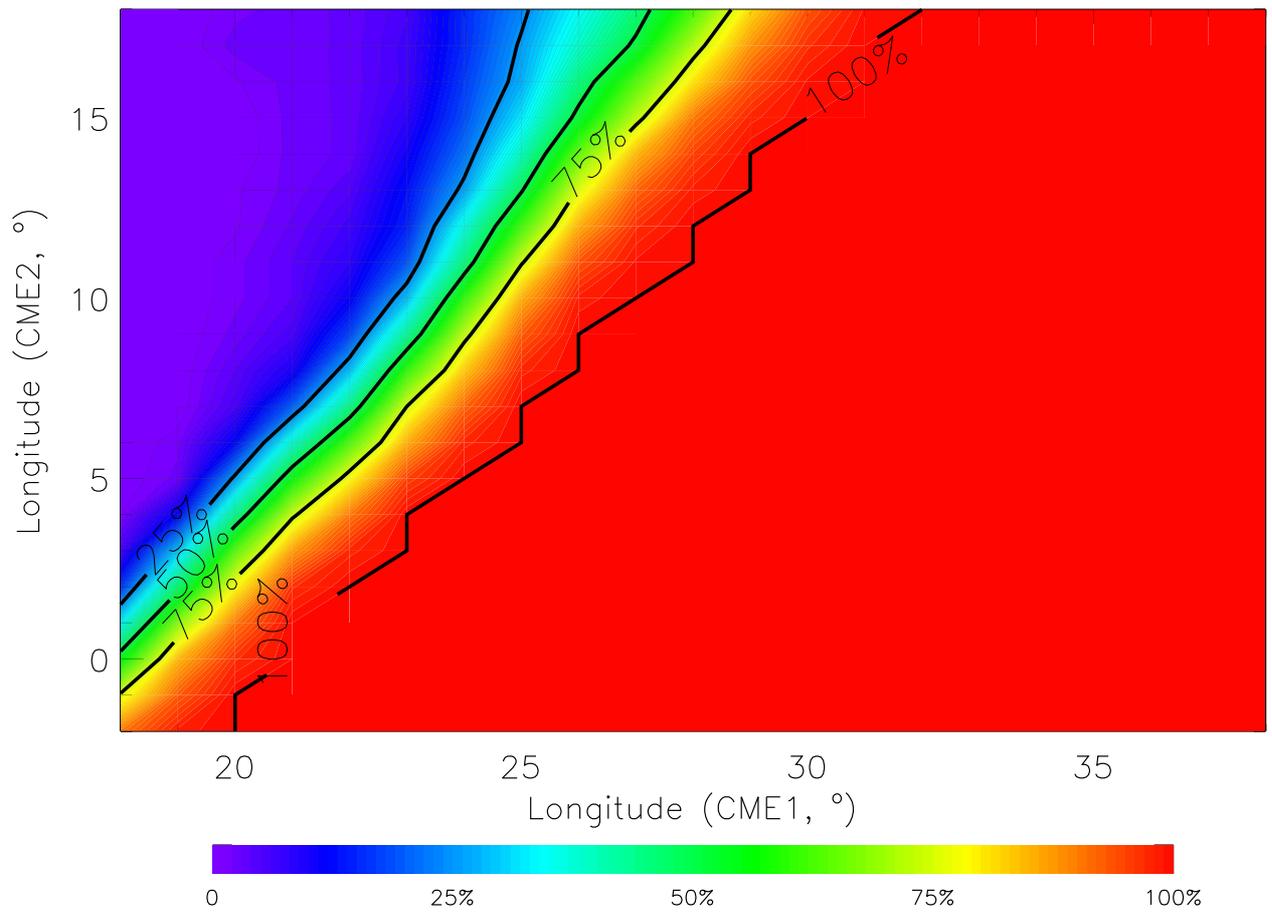}
\caption{Possibility of super-elastic collision. See the main text for details.}\label{long_vel_err}
\end{figure*}

The influence of large uncertainties, i.e., those in the CMEs'
longitudes and velocities as listed in Table \ref{table}, is further
examined. We sample the longitudes of the two CMEs at $1^\circ$
within the $10^\circ$ uncertainty. For each possible pair of
longitudes we consider an combination of five propagation speeds,
$[v_c\pm\Delta_{vc}, v_c\pm0.5\Delta_{vc}, v_c]$, for either of both
CMEs and five expansion speeds, $ [v_e\pm\Delta_{ve},
v_e\pm0.5\Delta_{ve}, v_e]$, for CME1, which constitute 125 cases.
Here, $\Delta_{vc}$ and $\Delta_{ve}$ are the uncertainties in the
CME speeds. For each case we are
able to obtain a value of $e$ and the change of the total kinetic
energy. The likelihood of super-elastic collision
for each longitude pair is therefore calculated.
Figure \ref{long_vel_err} presents the result.
Most area shows a strong likelihood of super-elastic
collision. Specifically, \mod{72.6\%} are more than 75\% likely, and \mod{63.0\%} are 100\% likely, to experience a
super-elastic collision. In contrast, as a few as \mod{6.3\%} combinations are definitely non-super-elastic. Overall,
it is \mod{72.8\%} likely for the collision to be super-elastic.

\section{Source of kinetic energy gain}
The source of the net kinetic energy gain and the way/condition of the energy
conversion are key issues for super-elastic collisions.
The divergent configuration of solar wind implies that the internal
pressure of a CME is always stronger than the external pressure when
it move away from the Sun, which is the main cause of the CME expansion.
In this process, the magnetic and thermal energies of the CME are
continuously dissipated\cite{Kumar_Rust_1996, Wang_etal_2009}.
It could be estimated that, for a typical CME at 1 AU with the magnetic
field strength of 10 nT, temperature of $10^5$ K, density of 5 cm$^{-3}$ and velocity of 500 km s$^{-1}$,
the magnetic and thermal energy density is about 6\% of the
kinetic energy density. The percentage will be much higher when the CME
is closer to the Sun. Thus, the magnetic and thermal energy of CMEs should
be sufficient to provide a $\sim\mod{6.6}$\% increase of kinetic energy in the
super-elastic collision, and the persistent expansion of CMEs may provide
the way for the magnetic/thermal energy to convert into kinetic energy.

Besides, the detailed interacting process may be also important in
determining the nature of collision. An anti-correlation between the
impact velocity and the coefficient of restitution was reported in
collisions among ice particles of Saturn B
ring\cite{Bridges_etal_1984} and granular
materials\cite{Hayakawa_Kuninaka_2004}. Experiments and simulations
on granular materials and nanoclusters have further shown that the
collision between a hard sphere and a soft plate tends to be
super-elastic\cite{Louge_Adams_2002, Kuninaka_Hayakawa_2004,
Saitoh_etal_2010}. These imply that super-elastic collision requires
sufficient interaction time and touching area for momentum exchange
and energy conversion. In our case, the compression and restitution
phases lasted about 16 hours, during which a clear arc-shaped
structure stayed visible for about 7 hours. These phenomena suggest
that the two CMEs had sufficient time and sufficiently large
touching area to convert magnetic/thermal energy into kinetic
energy. It is worthy of further investigation to see if a similar
anti-correlation applies to CME collisions, i.e., larger coefficient
of restitution corresponds lower impact velocity.

Although in granular physics, rotational motion and thermal
fluctuation have been considered the possible mechanism for the
increased kinetic energy\cite{Louge_Adams_2002,
Kuninaka_Hayakawa_2004, Kuninaka_Hayakawa_2009, Saitoh_etal_2010},
they are probably not suitable for CME collisions. First, there is
no evidence that plasma within a CME undergoes a significant
rotations in interplanetary space. Second, CMEs are a large-scale
structure with huge mass and thus the thermal fluctuation of
microscopic particles should not be able to affect the macroscopic
behavior of CMEs.

In summary, the good match between the predictions by the simplest
collision model and the observations suggests that such large-scale
magnetized plasmoids could be simplified as balls instead of using
complicated magnetohydrodyanmics or plasma kinetic theories in
studying their collision. The collision may be super-elastic,
through which the system gains kinetic energy from the
magnetic/thermal energy of CMEs. Of course, the process and
consequence might be different if significant reconnection occurs in
the collision region. This will be another issue.

\bibliographystyle{naturemag}
\bibliography{../../ahareference}

\acknowledgments{
We acknowledge the use of data from SECCHI, IMPACT, PLASTIC, WAVES instruments on STEREO,
LASCO on SOHO and WAVES on WIND. STEREO is the third mission in
NASA's Solar Terrestrial Probes program, and SOHO is a mission of
international cooperation between ESA and NASA. We thank anonymous
referees for their constructive comments. Y.W. also thanks No\'{e}
Lugaz for some valuable discussion. This work is supported by grants
from the 973 key project 2011CB811403, NSFC 41131065, 40904046,
40874075, and 41121003, CAS the Key Research Program KZZD-EW-01-4,
100-talent program, KZCX2-YW-QN511 and startup fund, and MOEC
20113402110001 and the fundamental research funds for the central
universities.

[Author Contributions] Y.W. designed the analysis of the collision and performed the
theoretical derivations. C.S. found this event and made the data processing and calculations.
S.W. gave constructive suggestions on the analysis of the collision. Y.L. gave some advice
on the construction of the Jmap and interpretation of the elongation angle. B.M. carried out
the literature investigation and provided valuable additions. A.V. calculated the masses of
the CMEs and gave many valuable suggestions. R.L. and P.Y. participated in
the discussion and gave many suggestive comments. J.L. and Z.Z. made a contribution to the
data analysis.

[Author Information] The authors declare no competing financial interests.
Correspondence and requests for materials should be addressed to Y.W. (ymwang@ustc.edu.cn).
}

\end{document}